\documentclass[runningheads]{llncs}
\usepackage[T1]{fontenc}
\usepackage{amsmath}
\usepackage{amssymb}
\usepackage{float}
\usepackage{subcaption}
\usepackage{graphicx}
\usepackage{adjustbox}
\usepackage{epigraph}
\usepackage{xcolor}
\newif\ifrevise
\revisetrue 

\newcommand{\blfootnote}[1]{%
  \begingroup
  \renewcommand{\thefootnote}{}\footnote{#1}%
  \addtocounter{footnote}{-1}%
  \endgroup
}

\begin{document}
\title{Optimization of Deep Learning Model for Denoising Dose Profiles in Proton Therapy}
\author{Junaid Jawaid\inst{1} \and
Matteo Spezialetti\inst{1} \and
Gabriele Libardi\inst{3}\and
Giovanni Stilo\inst{2} \and
Filippo Mignosi\inst{1}}
\authorrunning{Junaid Jawaid et al.}
\institute{
DISIM, University of L'Aquila, L'Aquila, Italy\\
\and
Department of AI, Data and Decision Sciences,
Luiss Guido Carli, Rome, Italy
\and
Marbl Energy, Vienna, Austria
}
\maketitle
\blfootnote{This is a preprint of a paper accepted for publication in the proceedings of FAIEMA 2026, to appear in \emph{Lecture Notes in Computer Science}, Springer. The final authenticated version, once published, will be available at Springer via the corresponding DOI; please cite that version where possible. This preprint has not undergone Springer's copyediting and typesetting and may differ from the version of record.}
\begin{abstract}
Proton therapy offers a dosimetric advantage over conventional radiotherapy
through the Bragg peak, enabling precise dose delivery to the tumour while
sparing surrounding tissue. Monte Carlo (MC) simulation provides the most
accurate dose calculation but is computationally expensive for clinical
workflows. A promising compromise is to simulate with a reduced number of
primary histories and recover a high-quality dose distribution through a
deep learning denoising step.

In this work, we take a 3D multiscale U-Net from recent literature as a
case study and investigate two improvements critical for clinical deployment.
First, we replace the standard MSE training objective with a differentiable
$\gamma$-index-based loss, extended to native 3D
operation, and demonstrate that directly optimising the clinical evaluation
metric yields a mean Gamma Passing Rate of 99.47\% at 2mm/2\% and 95.67\%
at 1mm/1\% on the test set, compared to 92.10\% and 71.13\% for MSE.
Second, we apply L1-norm structured filter pruning to reduce the model's memory footprint, achieving
a 13.3\% parameter reduction with negligible performance loss under partial
pruning, and 20\% reduction under full-block pruning after fine-tuning.
\end{abstract}

\section{Introduction}

Proton therapy has established itself as a major clinical modality in oncology,
offering a dosimetric advantage over conventional photon-based radiotherapy
through the characteristic Bragg peak, which allows precise energy deposition
at the tumour site while sparing surrounding healthy tissue. According to the
Particle Therapy Co-Operative Group (PTCOG), the number of treatment facilities
and patients treated worldwide has grown steadily, with over 220,000 patients
treated by 2019. Despite this clinical success, treatment planning for proton
therapy remains computationally demanding. As identified in a 2020 special issue
of The British Journal of Radiology \cite{nystrom2020treatment}, more accurate and faster
dose calculation algorithms remain an open challenge for the field.

The dominant clinical approach to dose calculation relies on Analytical Dose
Calculation (ADC), also known as the Pencil Beam (PB) algorithm, which offers
fast computation at the cost of reduced accuracy \cite{saini2018advanced}, particularly in thoracic
tumours where tissue heterogeneity is pronounced \cite{maes2018advanced,taylor2017pencil,teoh2020analytical}.
Monte Carlo (MC) simulation, by contrast, models particle transport with high
physical fidelity and is considered to be the gold standard for proton dose
calculation. Its clinical adoption is, however, limited by computational cost:
achieving statistically reliable dose distributions requires simulating an amount of
particle histories, which can be unviable on conventional
hardware. Several strategies have been explored to address this bottleneck,
including fast MC approximations and GPU-parallelised particle transport \cite{schiavi2017fred,yepes2010gpu}. Deep
learning offers a complementary and increasingly promising third approach,
capable of mapping noisy, low primary particle count MC simulations to high-quality
dose distributions at inference speeds of only a few seconds, which could be a
major step towards making online planning a reality.

In this direction, Spezialetti and co-authors \cite{Matteo_2020_unet_IEEE} proposed a
multiscale 3D convolutional network that takes as input a CT scan and a
$5\times10^3$ primary particle MC dose distribution and produces a denoised output
equivalent in quality to a $10^6$ primary particle simulation. The network, trained with
a standard Mean Squared Error (MSE) loss, demonstrated that deep learning is a
viable route to fast and accurate proton dose refinement. However, MSE is a
pixel-wise objective that does not directly optimise the metric of clinical
relevance in radiotherapy dose evaluation: the Gamma Passing Rate (GPR).
Martinot et al.\ \cite{Martinot2023diffGammaLoss} addressed this gap by
formulating a differentiable surrogate for the GPR, embedding the gamma index
within a parametric sigmoid to enable gradient-based optimisation. Their work,
however, was demonstrated on 2D dose arrays with bilinear interpolation.

Inspired by the possibility of taking advantage of both contributions above, we consider the network proposed in \cite{Matteo_2020_unet_IEEE} as our case study.

Specifically, we organise the study around three research questions:

\begin{itemize}
    \item \textbf{RQ1.} Does replacing the pixel-wise MSE objective
    with a differentiable surrogate of the Gamma Passing Rate , extended to native 3D operation paired with automatic mixed precision improve dose agreement under clinically relevant
    evaluation criteria, and does it improve the worst-case behaviour that
    matters for clinical acceptability, not only the mean?

    \item \textbf{RQ2.} To what extent can the resulting network be
    compressed by L1-norm structured filter pruning without degrading dose
    agreement, and how does that trade-off vary with the scope of pruning
    applied to the multiscale elaboration blocks, refer to figure \ref{fig:matteo_net}?

    \item \textbf{RQ3.} Does the parameter reduction obtained by
    pruning translate into a corresponding reduction in inference latency, and
    if not, what does that imply for the deployment of models of this class?
\end{itemize}

\noindent In addressing these questions we make the
following contributions:

\begin{itemize}
    \item We extend the differentiable gamma index loss of Martinot et al.\
    \cite{Martinot2023diffGammaLoss} to native 3D operation using trilinear
    interpolation in PyTorch, and apply it to the proton dose denoising task, to the best of our knowledge,
    for the first time in a fully volumetric setting.

    \item We adopt \texttt{bfloat16} automatic mixed precision (AMP) training to
    reduce GPU memory consumption.

    \item We apply L1-norm based structured filter pruning
    \cite{Li_2017_PruningFilters} to the MultiScaleElaboration blocks of the
    network under two scopes of increasing aggressiveness: partial pruning of
    the deeper convolutional layers
    and full pruning of all seven convolutions per block. We report the trade-off
    between model compression and GPR under both configurations.

    \item We provide a sensitivity analysis of accuracy and memory metrics as a
    function of pruning scope, and demonstrate that partial pruning achieves a
    13.3\% parameter reduction with a GPR drop of only 0.06\% at
    2mm/2\% and 0.64\% at 1mm/1\%, while full-block
    pruning achieves 20\% reduction which, after fine-tuning, retains
    clinical dose agreement at 2mm/2\% (99.33\%) but incurs a residual 1.22\%
    loss at the stricter 1mm/1\% criterion.The resulting compressed models
    retain clinical dose agreement and therefore provide a cheaper backbone for
    subsequent experimentation on this task.
\end{itemize}
\section{Related Work}

\subsection{Deep Learning for Proton Dose Denoising and Prediction}

The U-Net architecture \cite{First_unet}, originally proposed for biomedical
image segmentation, has proven well-suited to dense volumetric prediction tasks
in radiotherapy. Nguyen et al.\ \cite{nguyen2019hdUnet} introduced a
Hierarchically Densely Connected U-Net for 3D dose distribution prediction in
head and neck radiotherapy, demonstrating that dense intra-level connections
achieve comparable accuracy with 12 times fewer trainable parameters than a
standard U-Net. Wu et al.\ \cite{wu2021improving} further showed that a
U-Net-based model can convert pencil beam dose distributions to Monte Carlo
accuracy across multiple tumour sites, achieving average gamma passing rates of
92--99\% at the 1mm/1\% criterion with sub-4-second inference per field. The
architecture central to this study is that proposed by Spezialetti et al.\
\cite{Matteo_2020_unet_IEEE}, which processes parallelepiped volumes of shape
$256\times32\times32$ from a two-channel input (CT scan, and $5\times10^3$
particle MC dose) and was trained with a standard MSE loss. The memory requirement of this architecture can be high, and it grows cubically with the edge length of the input volume. Our work builds
directly on this network, augmenting its training objective and reducing its
memory footprint.

\subsection{Loss Functions and the Gamma Index in Dose Prediction}

Standard voxel-wise losses such as MSE are the most common training objectives
in radiotherapy dose prediction \cite{nndosenet2025}, but they are not aligned
with the clinical evaluation metric of interest, namely the Gamma Passing Rate
(GPR). The gamma index \cite{Low1998gamma} (equation \eqref{eq:gamma_index}), defined as the minimum generalised
distance in combined dose-difference and spatial space between a reference and
evaluated distribution, provides a joint assessment of dose accuracy and
distance-to-agreement. Sharp loss
\cite{sharploss2021} demonstrated statistically significant improvements over
MSE in mean absolute dose differences across planning target volumes and organs
at risk. Most directly relevant to our contribution, Martinot et al.\
\cite{Martinot2023diffGammaLoss} formulated a differentiable surrogate for the
GPR by embedding the gamma index within a parametric sigmoid, enabling gradient
flow through an otherwise discontinuous pass/fail criterion. Their study
demonstrated improved denoising performance when training MC dose simulation
networks with this loss; however, their implementation operated on 2D arrays
with bilinear interpolation. In this work we extend the formulation to native 3D volumes using trilinear
interpolation in PyTorch, and apply it as a standalone training objective to
the Spezialetti et al.\ network \cite{Matteo_2020_unet_IEEE}, comparing it
directly against MSE as a controlled baseline.

\subsection{Model Compression via Structured Pruning}

Filter pruning, which removes entire convolutional filters ranked by a
structural importance criterion, produces physically smaller dense networks
requiring no sparse-tensor runtime support. Li et al.\
\cite{Li_2017_PruningFilters} established the L1-norm of filter weights as an
effective and computationally inexpensive ranking criterion, forming the basis
of the pruning strategy adopted in this work. In the medical imaging domain,
Adnan et al.\ \cite{Adnan_2024_StructuredPruningPathology} demonstrated that
structured pruning of U-Net-style architectures achieves compression of 70\%
or more with negligible loss in segmentation performance, and Sreeram and Kainz
\cite{sreeram2023sculpting} showed that aggressive filter pruning of a cardiac
segmentation model yielded over $1000\times$ compression with only a $\sim$4\%
quality reduction. Consistent with the conservative accuracy-preservation
strategy recommended for clinical models
\cite{Choudhary_2021_BreastCancerPruning,RamosCortez_2025_LightweightUNet},
we validate pruning on held-out batches prior to committing weight removal, and
subsequently fine-tune the pruned network to recover any performance loss.
\section{Materials and Methods}

This study investigates two aspects of the denoising pipeline of Spezialetti et
al.\ \cite{Matteo_2020_unet_IEEE}: the training objective and the compressibility
of the trained model. The same architecture is trained from scratch under
identical conditions with two losses, a standard MSE baseline replicating the
original setup and the differentiable $\gamma$-index loss of Martinot et al.\
\cite{Martinot2023diffGammaLoss} extended here to native 3D, both with automatic mixed precision. Both models are
evaluated on a held-out test set with the clinical GPR metric
(\texttt{pymedphys}) under 2mm/2\% and 1mm/1\% criteria, and structured filter
pruning is then applied to the better-performing GPR model to assess the
compression--accuracy trade-off. The following subsections describe the dataset,
architecture, loss, post-processing, mixed precision training, and pruning.
\subsection{Dataset}
The dataset used in this study follows the construction procedure described in
detail by Spezialetti et al.\ \cite{Matteo_2020_unet_IEEE}, to which we refer
the reader for full methodological details and visual illustrations. Briefly, a
subset of 39 CT studies was selected from a collection of 105 patients imaged
via Computed Radiography, Direct Digital Capture, and CT. All volumes were
interpolated to a uniform resolution of $256\times256\times256$ cubic voxels
with a side length of 2\,mm, and reoriented so that the $x$, $y$, and $z$ axes
correspond to the left--right, bottom--up, and posterior--anterior patient
directions respectively.

Crucially, and following the original protocol, the 39 CT studies were
first partitioned at the patient level into disjoint training, validation, and
test sets. All samples derived from a given study, and therefore from a
given patient, reside entirely within a single split; no rotated or
translated variant of a training patient can appear in the validation or test
set. This precludes patient-level data leakage between the sets.

Multiple samples were generated from each volume through a combination of
rotations and translations to maximise dataset size while limiting overlap
between samples. For the training and validation sets, 13 equispaced rotations
around the $y$-axis (from $-90°$ to $+90°$) were applied, each followed by a
translation along $y$ in steps of 8 pixels, and additionally combined with
alternating $\pm15°$ rotations around $x$. For the test set, a reduced set of
rotations was applied to maximise independence between samples. From each
resulting transformed volume, the $32\times32\times256$ subvolume centred on
the $z$-axis was extracted, yielding the parallelepiped geometry on which the
network operates. Beams were simulated along the negative $z$-axis direction to
avoid interaction with the patient bed structure.

Monte Carlo dose simulations were performed using TOPAS \cite{topas_validation}.
A pencil proton beam source was placed at coordinates $(0\,\text{m},
0\,\text{m}, 2\,\text{m})$ directed toward the world origin. Beam energies were
sampled uniformly from the range $[87, 162]$\,MeV, corresponding to Bragg peak
depths in water of approximately 6--18\,cm. Beam spot sizes were drawn from
$\sigma \in \{2, 3, 5\}$\,mm. Hounsfield Unit-to-material conversions followed
the procedure of Schneider et al.\ \cite{schneider2000}. Each geometry was
simulated at both $5\times10^3$ and $10^6$ primary histories. The low primary particle count simulation serves
as the noisy network input, and the high primary particle count simulation as the denoising
target, and both were normalized to $10^5$ particles.

To further augment the training and validation sets, artificially perturbed
patient geometries were generated by introducing random density perturbations to
5\% of non-air voxels (HU $> -950$), smoothed with a 3D Gaussian filter
($\sigma = 0.5$), to encourage the network to generalise to anatomies with
abrupt density transitions. Following this procedure, the final dataset
comprised 624 training samples, 208 validation samples, and 266 test samples. During training, validation, and testing, rotational data augmentation was applied along the elongated axis.

\subsection{The Network}

The network used in this study is based on the U-Net architecture
\cite{First_unet}, whose encoder--decoder structure with skip connections is
well-suited to dense volumetric prediction and has been widely adopted for dose
estimation in both photon and proton therapy
\cite{nguyen2019feasibility,wu2021improving}. The encoder compresses the input
into a compact latent representation and the decoder reconstructs the output,
while skip connections carry spatially detailed features to the decoder and
provide shortcut paths for gradient flow.

The specific architecture used in this study was proposed by Spezialetti et
al.\ \cite{Matteo_2020_unet_IEEE} and is inspired by the 3D HD U-Net
\cite{nguyen2019hdUnet}, itself a fusion of U-Net \cite{First_unet} and
DenseNet. Beyond the HD U-Net foundation, the architecture further incorporates
Inception-like multiscale elaboration modules, drawing inspiration from
Inception-ResNet and InceptionV4, which process features at multiple spatial
scales in parallel before concatenation. This design reflects three guiding
principles: a reduction-expansion structure to combine features at multiple
levels of abstraction, Inception-like modules to capture volumetric variations
at different scales simultaneously, and DenseNet-style layer concatenations to
maintain direct feature access across layers. Additionally, shortcut
concatenations provide ResNet-like gradient propagation advantages, improving
training stability. The dense convolutional layers of the original HD U-Net were
replaced with these Inception-inspired multiscale blocks, short-cut through
concatenations with their inputs.

Input volumes
have shape $256\times32\times32$, reflecting the parallelepiped subvolumes
extracted along the beam direction. To handle this asymmetric shape, the network
performs three steps of reduction and expansion. A $1\times1\times1$ convolution
is applied before each expansion block to compress the feature map, roughly
doubling training speed and reducing GPU memory consumption, a necessary
trade-off for 3D volumetric networks where memory pressure is substantial. The
network accepts a two-channel input consisting of the CT scan and the
$5\times10^3$ primary particle MC dose distribution, and produces a single-channel
denoised dose output equivalent in quality to a $10^6$ primary particle simulation. The
original model was trained with MSE loss using TensorFlow. In this study, we
reimplemented the architecture in PyTorch and trained it from scratch under two
separate objectives, MSE and the differentiable GPR loss
\cite{Martinot2023diffGammaLoss} extended to 3D, both done with bfloat automatic mixed precision, to evaluate whether
directly optimising the clinical evaluation metric improves denoising
performance. Structured pruning is subsequently applied to the
best-performing model to further reduce its memory footprint.
\begin{figure}[htbp]
\includegraphics[width=\textwidth]{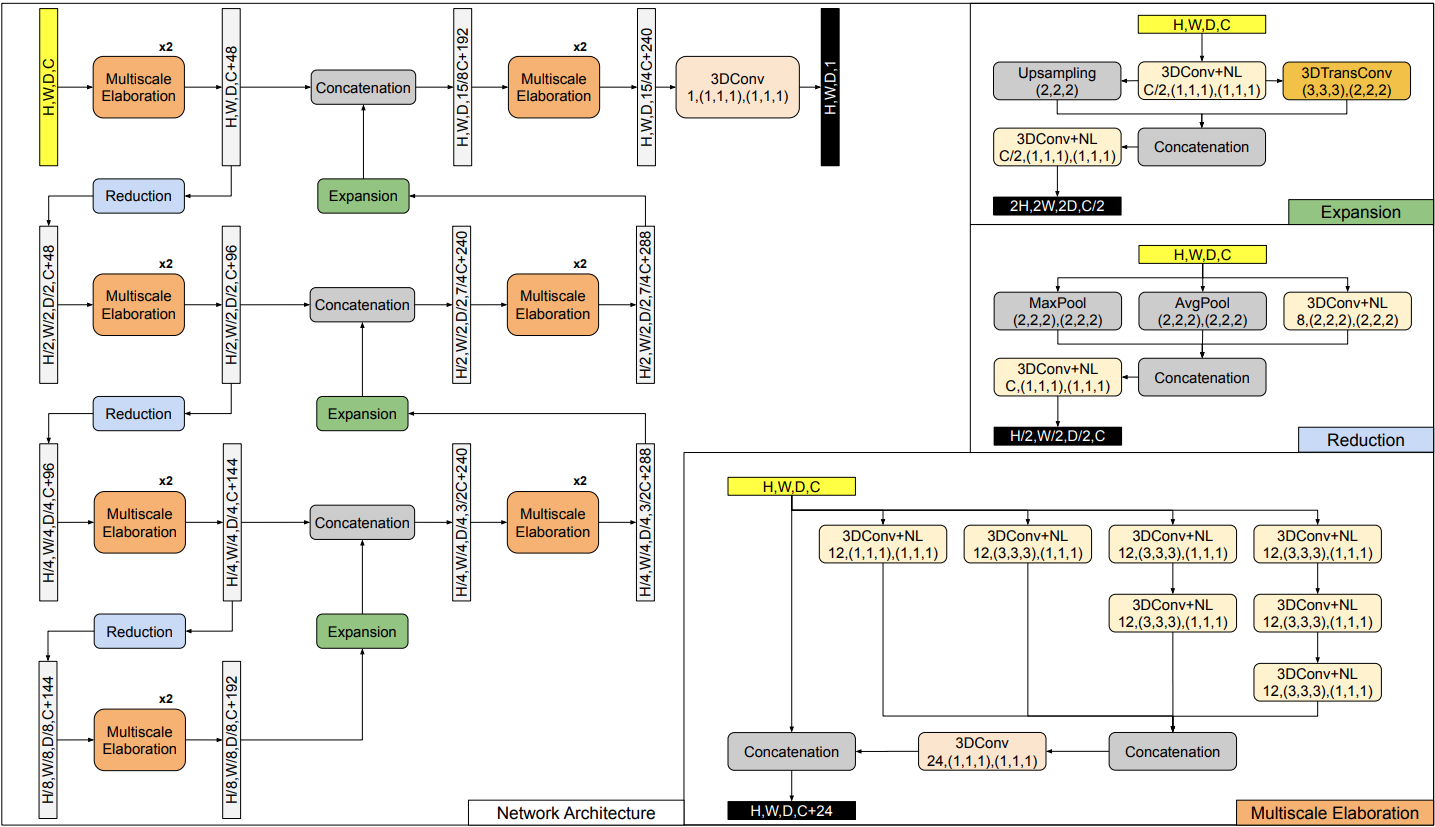}
\caption{Network architecture, taken from Spezialetti et al.\ \cite{Matteo_2020_unet_IEEE}.} \label{fig:matteo_net}
\end{figure}
\subsection{Gamma Loss Function}
Conventional voxel-wise losses such as MSE do not directly optimise the Gamma
Passing Rate (GPR), the clinical evaluation metric of interest. GPR is itself
non-differentiable, but it can be approximated by the differentiable surrogate of
\cite{Martinot2023diffGammaLoss}, in which the gamma index is embedded within a
parametric sigmoid to construct a smooth objective.

The Gamma Index $\Gamma$ at a reference voxel position $\mathbf{r}$ is defined as the minimum generalized distance from the ground-truth dose $y(\mathbf{r})$ to any point $\mathbf{r} + \delta$ in the predicted dose distribution $\hat{y}$. Given a dose difference threshold $\Delta$ and a Distance-to-Agreement threshold $DTA$, the local gamma value is calculated as:

\begin{equation}
\Gamma(\mathbf{r}) = \min_{\delta \in \mathcal{S}} \sqrt{ \frac{|\delta|^2}{DTA^2} + \frac{\left(\hat{y}(\mathbf{r}+\delta) - y(\mathbf{r})\right)^2}{\Delta^2} }
\label{eq:gamma_index}
\end{equation}
where $\delta$ represents the spatial offset within a search neighbourhood $\mathcal{S}$ defined by $|\delta| \leq DTA$. To ensure differentiability, a soft pass/fail criterion is introduced using a parametric sigmoid function controlled by a sharpness parameter $\beta$. We define the parametric sigmoid generically as

\begin{equation}
\sigma_{\beta}(x) = \frac{1}{1 + e^{-\beta x}}
\label{eq:sigmoid}
\end{equation}

This formulation acts as a differentiable approximation of the standard Heaviside step function. For all points $\mathbf{r} \in \Omega$, the sigmoid converges to the hard pass/fail criterion as the sharpness parameter increases:

\begin{equation}
\lim_{\beta \to +\infty} \sigma_{\beta}(1 - \Gamma(\mathbf{r})) = \mathbb{1}_{{\Gamma(\mathbf{r}) \leq 1}}
\label{eq:limit}
\end{equation}

The final loss function is obtained by computing the Soft Gamma Pass Rate (Soft-GPR) over a region of interest defined by a dose cutoff threshold $\tau$:

\begin{equation}
\mathcal{L}_{GPR}(\hat{y}, y) = 1 - \frac{\sum_{\mathbf{r} \in \Omega} \sigma_{\beta}(1 - \Gamma(\mathbf{r}))\cdot \mathbb{1}_{{y(\mathbf{r}) \geq \tau}}}{\sum_{\mathbf{r} \in \Omega} \mathbb{1}_{{y(\mathbf{r}) \geq \tau}}}
\label{eq:gpr_loss}
\end{equation}

\begin{description}
\item[$\Omega$:] set of all voxel positions in the 3D spatial domain
\item[$\mathbf{r}$:] voxel position, $\mathbf{r} \in \Omega$
\item[$y(\mathbf{r})$:] ground-truth dose at voxel $\mathbf{r}$
\item[$\hat{y}(\mathbf{r})$:] predicted dose at voxel $\mathbf{r}$
\item[$\tau$:] dose cutoff threshold defining the region of interest
\item[$\beta$:] sharpness parameter controlling the smoothness of the sigmoid surrogate
\end{description}

Minimizing $\mathcal{L}_{GPR}$ therefore corresponds to maximizing a differentiable approximation of the Gamma Passing Rate. During training, $\beta$ is typically annealed to gradually transition from smoother gradients toward a sharper approximation of the binary pass/fail boundary.

\subsubsection{Three-dimensional search and interpolation.}
The formulation above is evaluated natively in 3D. Both the predicted
and reference volumes are first resampled from the native $2\,\text{mm}$ isotropic
grid to a $1\,\text{mm}$ grid by trilinear interpolation
(\texttt{torch.nn.functional.interpolate}). This refines the resolution at which
the distance-to-agreement term is resolved and keeps the operation
differentiable

The minimisation in Eq.~\eqref{eq:gamma_index} runs over a discrete
neighbourhood $\mathcal{S}$ of offsets $\delta$ with $|\delta| \leq DTA$ on the
resampled grid, forming a sphere about each reference voxel. For every offset the
combined dose-difference and distance term is evaluated across the whole volume
in parallel and the elementwise minimum yields $\Gamma(\mathbf{r})$; the search
is thus a dense GPU sweep over shifted copies of the predicted volume rather than
a per-voxel iterative search, and gradients propagate through the minimising
offset at each voxel. Gamma values are clamped at $\max\Gamma = 10$ to bound
voxels with no agreement in the neighbourhood and prevent outliers from
dominating the gradient.

In our experiments the gamma evaluation criteria during training were
set to $DTA = 2\,\text{mm}$ and $\Delta = 2\%$, with a lower dose cutoff
threshold $\tau = 20\%$. Note that this resampling and search procedure is
applied only when computing the training loss; the reported test-set GPR values
are computed independently with \texttt{pymedphys}, which performs its own
interpolation internally, so that evaluation does not depend on our loss
implementation.

\subsection{Mixed Precision Training}

To reduce GPU memory consumption during training on $2\,\text{mm}^3$ resolution
3D dose volumes, we adopt Automatic Mixed Precision (AMP) \cite{micikevicius2018amp}
via PyTorch's \texttt{torch.amp} autocast. We use \texttt{bfloat16} as
the low-precision compute type: compute-intensive operations such as convolutions
are executed in \texttt{bfloat16}, while numerically sensitive reductions and the
FP32 master weights are retained at full precision. Unlike \texttt{float16},
\texttt{bfloat16} preserves the full 8-bit exponent, and therefore the dynamic
range of FP32, differing only in mantissa precision; gradients consequently do
not underflow, and no loss scaling (\texttt{GradScaler}) is required for stable
convergence. Operating on 16-bit tensors halves the memory footprint of the
low-precision activations and weights relative to FP32 and yields 1.5--2$\times$
throughput improvements on Tensor Core hardware \cite{NVIDIA_DLPerf_2023}. The
choice of 8 as the pruning target channel count is also motivated by this: NVIDIA
cuDNN requires channel counts to be multiples of 8 to trigger Tensor Core
acceleration in FP16/BF16 mode
\cite{NVIDIA_cuDNN_2023}\cite{Markidis_2018_TensorCore}, meaning the original
12-channel elaboration blocks were incompatible with this acceleration path
entirely.

\subsection{Post-Processing: Dilated Dose Mask}\label{sec:mask}

Training exclusively with $\mathcal{L}_{GPR}$ produced dose distributions with
strong agreement in the clinically relevant high-dose region, but introduced
spurious non-zero activations in voxels where the true dose is zero or
negligibly small, as visible in Figure~\ref{fig:qualitative} where a faint
signal appears in regions that should be zero. This artefact arises directly
from the structure of the GPR loss: as defined in Eq.~\eqref{eq:gpr_loss},
$\mathcal{L}_{GPR}$ only evaluates voxels where the reference dose
$y(\mathbf{r})$ exceeds the cutoff threshold $\tau$, leaving the network's
predictions in the low-dose and zero-dose regions entirely unsupervised during
training. MSE, by contrast, penalises all voxels uniformly, implicitly
suppressing out-of-field activations. The result is that a network trained
solely with $\mathcal{L}_{GPR}$ is free to produce arbitrary values outside
the irradiated volume, which manifests as visible noise artefacts in regions
that should be zero.

To suppress these artefacts without modifying the loss function or
reintroducing MSE, we applied a post-processing mask derived from the
$5\times10^3$ primary particle dose $x_{\text{MC}}$, which is already available as
the second input channel to the network. The mask is computed in three steps.
First, $x_{\text{MC}}$ is binarised at $\tau_{\text{mask}} = 10\%$ of its
maximum value to produce an initial region-of-interest mask. Second, this
mask is morphologically dilated in 3D using a spherical structuring element
of radius 5 voxels, to account for the spatial spread introduced by the
denoising network and to avoid clipping the Bragg peak penumbra. Third, the
predicted dose is zeroed outside the dilated mask. The final masked output
$\hat{y}_{\text{masked}}$ is:

\begin{equation}
    \hat{y}_{\text{masked}}(\mathbf{r}) = \hat{y}(\mathbf{r}) \cdot M(\mathbf{r})
\end{equation}

\noindent where the binary mask $M$ is defined as:

\begin{equation}
    M(\mathbf{r}) = \text{dilate}\!\left(
        \mathbb{1}_{x_{\text{MC}}(\mathbf{r}) \geq \tau_{\text{mask}}}
    \right), \quad \tau_{\text{mask}} = 0.1 \cdot \max(x_{\text{MC}})
\end{equation}

\noindent This approach is self-contained, requiring no additional input
beyond what the network already receives, and adds negligible computational
overhead at inference time since the mask computation involves only a single
thresholding and a morphological dilation on a $256\times32\times32$ volume.

The reported GPR is invariant to this mask by construction, so it
confers no metric advantage. \texttt{pymedphys} computes the GPR only at
reference voxels above the $20\%$ cutoff, whereas the mask zeroes predicted dose
solely \emph{outside} the dilated region; since that region is seeded at the
lower $10\%$ input threshold and dilated by a $5$-voxel spherical element, it
comfortably encloses the $\geq\!20\%$ voxels the metric evaluates. No voxel
entering the GPR summation is therefore modified, and the MSE and GPR models are
compared on an equal footing over the same above-cutoff voxels. The mask serves
only to suppress out-of-field artifacts and produce clean dose maps.

\subsection{Pruning}

To reduce the memory footprint of the trained 3D U-Net and to assess
whether its parameter count could be reduced without compromising dosimetric
accuracy, we applied structured filter pruning to the MultiScaleElaboration
blocks of the GPR-trained model. Following the criterion established by
\cite{Li_2017_PruningFilters}, filters were ranked by their L1-norm, the
sum of absolute kernel weights, and the lowest-ranked filters were removed
alongside their corresponding instance normalisation parameters and downstream
input channels, producing a physically smaller dense network that requires no
sparse-tensor runtime support. The target channel count after pruning was
aligned to the nearest multiple of 8, reducing the internal elaboration channels
from 12 to 8, the lowest width compatible with the Tensor Core alignment
constraint discussed above (the original 12-channel width, not a multiple of 8,
precludes that acceleration path \cite{NVIDIA_cuDNN_2023,Markidis_2018_TensorCore}).

To protect architecturally sensitive connections, specifically the three
conv--IN--conv sequential pairs within each MultiScaleElaboration block,
pruning was applied pairwise: the output channels of the upstream convolution
and instance normalisation layer were sliced simultaneously with the
corresponding input channels of the downstream convolution, preserving the
dimensional consistency of each residual path without requiring an external
dependency-graph library \cite{Adnan_2024_StructuredPruningPathology}. Prior
to applying pruning to the full model, a dry-run was performed on a deep copy
using 20 validation batches; pruning was only committed if the resulting GPR
drop remained below 50\%, following the conservative accuracy-preservation
strategy recommended for clinical segmentation models
\cite{Choudhary_2021_BreastCancerPruning}\cite{RamosCortez_2025_LightweightUNet}.

We evaluated two pruning scopes of increasing aggressiveness. In the first
configuration, pruning was applied only to the deeper convolutional layers
within each block (conv3, conv4, conv4\_1) which we refer to as the \emph{partial MS pruned} net, leaving the
shallower branches intact. In the second configuration, pruning was extended
to all seven convolutions within every MultiScaleElaboration block across the
encoder, bottleneck, and decoder which we refer to as the \emph{full MS pruned} net. In both cases, the pruned model was
subsequently fine-tuned for up to 20 epochs using the Adam optimiser with a
learning rate of $10^{-6}$, $\beta = 4.45$ fixed during training as it
represented more closely the step pass/fail criterion, and a
ReduceLROnPlateau scheduler with patience of 3 epochs. Early stopping was
applied if no improvement was observed for 5 consecutive epochs, and the best
checkpoint was selected by validation GPR.

\subsection{Evaluation Metrics}
All models are evaluated with the Gamma Passing Rate (GPR), computed
using the \texttt{pymedphys} gamma index \cite{Low1998gamma}. The gamma index is
the standard metric for dose comparison in radiotherapy QA because it jointly
scores dose difference and distance-to-agreement. Neither alone suffices: a
dose-difference metric penalises a merely shifted distribution as harshly as a
wrong one, while a geometric metric ignores dose magnitude. This matters most in
steep-gradient regions such as the distal Bragg peak, where a sub-millimetre
shift yields a large local dose difference that a voxel-wise metric flags as a
severe error even when the distribution is clinically equivalent.

We report GPR at two criteria, $2\,\text{mm}/2\%$ and $1\,\text{mm}/1\%$,
both with a $20\%$ lower dose cutoff. The former is the routine clinical QA
criterion; the latter is stricter and exposes residual high-frequency error that
the clinical criterion saturates. The cutoff excludes near-zero out-of-field
voxels, where agreement is not clinically meaningful. We report the mean,
standard deviation, minimum and maximum, since the worst case matters more
clinically than the average. Inference time (forward pass, batch size~1) is a
secondary metric for RQ3.

\subsection{Training Setup}

All experiments were conducted on a workstation equipped with NVIDIA RTX
A4500 GPU, with all training performed on a single GPU with 20\,GB of VRAM.
All training runs used automatic mixed precision with \texttt{bfloat16} autocast
via PyTorch's \texttt{torch.amp} module \cite{micikevicius2018amp}, reducing
GPU memory consumption and enabling training on full $256\times32\times32$
volumes without memory overflow.

\subsubsection{Data Normalisation.}
All normalisation was applied in the data pipeline prior to being passed to the
network. CT volumes were linearly scaled as $(CT + 1000) / 3000$ as Hounsfield Unit of air is -1000 and the max value for the bone is 3000, mapping the
values in the range to approximately $[0, 1]$. The
$5\times10^3$ primary particle and $10^6$ primary particle dose volumes were each independently
min-max normalised to $[0, 1]$. Online data augmentation was applied by
rotating each sample by $0°$, $90°$, $180°$, and $270°$ around the axial plane
at load time, yielding four variants per stored sample. The network output was
max-normalised before being passed to the loss function:
\begin{equation}
    \hat{y}_{\text{norm}} = \frac{\hat{y}}{\max(\hat{y}) + \epsilon}
\end{equation}
\noindent where $\epsilon = 10^{-10}$ prevents division by zero.

\subsubsection{MSE Model.}
The MSE model was trained from scratch for up to 150 epochs using the Adam
optimiser with an initial learning rate of $10^{-4}$. A ReduceLROnPlateau
scheduler with a reduction factor of 0.7 and patience of 10 epochs was used to
decay the learning rate when validation loss plateaued. The batch size was 4
for training and for validation. The loss criterion was standard MSE computed
between the max-normalised network output and the min-max normalised
$10^6$ primary particle target dose. The best model checkpoint was selected by minimum
validation loss.

\subsubsection{GPR Model.}
The GPR model was trained from scratch for up to 150 epochs using the Adam
optimiser with an initial learning rate of $10^{-4}$, a ReduceLROnPlateau
scheduler with reduction factor 0.7, patience of 10 epochs, and a minimum
learning rate of $10^{-6}$. The batch size was 4 for both training and
validation. The loss criterion was $\mathcal{L}_{GPR}$ as defined in
Eq.~\eqref{eq:gpr_loss}, with dose difference threshold $\Delta = 2\%$ of the global maximum
dose in the target volume, distance-to-agreement $DTA = 2\,\text{mm}$, dose cutoff $\tau = 20\%$, and $\max\Gamma = 10$.
The dose difference threshold was applied globally, i.e.\ $\Delta =
0.02 \cdot \max(y)$, rather than as a local per-voxel percentage. Before the calculation of the gamma index, the target and predicted doses were scaled to $1\,mm^3$ resolution using `trilinear' interpolation.

The sharpness parameter $\beta$ was annealed with a two-phase linear schedule:
held at 0.1 during a warmup (epochs 1--20), raised to 2.0 over after epochs 20--65,
then to 5.0 over epochs 65--120 and held there, all updated linearly every 5 epochs. This
progressively transitions the soft-GPR surrogate from a smooth landscape toward
the sharp binary pass/fail boundary, improving early gradient signal while
converging toward the true metric. The best checkpoint was selected by maximum
validation GPR.

\section{Results and Discussion} \label{sec:results}

\begin{figure}[htbp]
    \centering
    \begin{subfigure}[t]{0.19\linewidth}
        \centering
        \adjustbox{trim={0.4\width} {0.4\height} {0.05\width} {0\height}, clip,
            width=\linewidth}{%
            \includegraphics{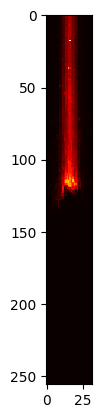}}
        \caption{}
        \label{fig:baseline}
    \end{subfigure}
    \hfill
    \begin{subfigure}[t]{0.19\linewidth}
        \centering
        \adjustbox{trim={0.4\width} {0.4\height} {0.05\width} {0\height}, clip,
            width=\linewidth}{%
            \includegraphics{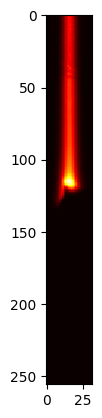}}
        \caption{}
        \label{fig:gt}
    \end{subfigure}
    \hfill
    \begin{subfigure}[t]{0.19\linewidth}
        \centering
        \adjustbox{trim={0.4\width} {0.4\height} {0.05\width} {0\height}, clip,
            width=\linewidth}{%
            \includegraphics{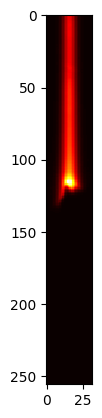}}
        \caption{}
        \label{fig:mse}
    \end{subfigure}
    \hfill
    \begin{subfigure}[t]{0.19\linewidth}
        \centering
        \adjustbox{trim={0.4\width} {0.4\height} {0.05\width} {0\height}, clip,
            width=\linewidth}{%
            \includegraphics{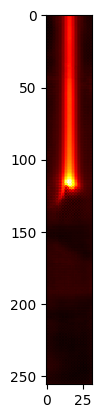}}
        \caption{}
        \label{fig:gpr}
    \end{subfigure}
    \hfill
    \begin{subfigure}[t]{0.19\linewidth}
        \centering
        \adjustbox{trim={0.4\width} {0.4\height} {0.05\width} {0\height}, clip,
            width=\linewidth}{%
            \includegraphics{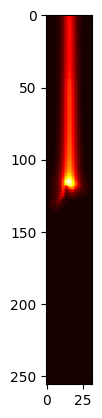}}
        \caption{}
        \label{fig:masked}
    \end{subfigure}
    \caption{Qualitative comparison of dose distributions for a representative
    test case: (a) noisy $5\times10^3$ primary particle input, (b) $10^6$ primary
    particle ground truth, (c) MSE prediction, (d) GPR prediction, showing spurious
    activations in zero-dose regions, and (e) GPR prediction with the dilated dose
    mask applied. The GPR model agrees more closely with the reference (b) in the
    high-dose region. For this case the MSE model scores 91.64\% / 69.96\% (2mm/2\%
    and 1mm/1\%) against 99.63\% / 97.28\% for the GPR model. A 20\% lower cutoff is
    used throughout and all panels are max-normalised.}
    \label{fig:qualitative}
\end{figure}

In this study, two models were trained from scratch on the architecture
depicted in Figure~\ref{fig:matteo_net}: one with MSE loss replicating the
original training setup of Spezialetti et al.\ \cite{Matteo_2020_unet_IEEE} with addition of mixed precision,
and one with the differentiable GPR loss extended to 3D. Both models were
evaluated on a held-out test set of 266 samples. Structured pruning was
subsequently applied to the GPR-trained model under two configurations.
All models were evaluated using the \texttt{pymedphys} gamma index
implementation under both 2mm/2\% and 1mm/1\% criteria with a 20\% lower
dose cutoff, which is consistent with clinical quality assurance practice.
Figure~\ref{fig:qualitative} shows a qualitative comparison of dose
distributions for a representative test case across all configurations,
and Table~\ref{tab:results} reports the full quantitative results.

Table~\ref{tab:results} summarises the Gamma Passing Rate (GPR) and inference
time across all model configurations, evaluated on the test set of 266 samples
using the \texttt{pymedphys} gamma implementation with a 20\% lower dose
cutoff.

\begin{table}[h]
\caption{\textbf{GPR and inference time results across all model configurations.}
Inference time measured as forward pass only at batch size 1.}
\label{tab:results}
\centering
\resizebox{\textwidth}{!}{%
\begin{tabular}{llccccc}
\hline
\textbf{Model} & \textbf{Criteria} & \textbf{Mean GPR (\%)} & \textbf{Std (\%)} & \textbf{Min (\%)} & \textbf{Max (\%)} & \textbf{Inference (ms)} \\
\hline
Baseline ($5\times10^3$) & 2mm/2\% & 26.75 & 16.66 & 6.52  & 73.09  & --- \\
MSE Only                 & 2mm/2\% & 92.10 & 8.01  & 43.32 & 99.97  & $92.22 \pm 4.90$ \\
Full Net (GPR Loss)      & 2mm/2\% & 99.47 & 1.12  & 85.86 & 100.00 & $91.53 \pm 15.35$ \\
Partial MS Pruned        & 2mm/2\% & 99.41 & 1.23  & 80.09 & 100.00 & $95.42 \pm 25.23$ \\
Full MS Pruned           & 2mm/2\% & 99.33 & 1.50  & 77.69 & 100.00 & $96.37 \pm 27.18$ \\
\hline
Baseline ($5\times10^3$) & 1mm/1\% & 13.54 & 8.67  & 2.48  & 37.65  & --- \\
MSE Only                 & 1mm/1\% & 71.13 & 20.23 & 7.42  & 99.27  & $89.16 \pm 3.02$ \\
Full Net (GPR Loss)      & 1mm/1\% & 95.67 & 6.70  & 34.81 & 99.96  & $91.78 \pm 15.18$ \\
Partial MS Pruned        & 1mm/1\% & 95.03 & 7.07  & 29.18 & 99.96  & $94.27 \pm 24.40$ \\
Full MS Pruned           & 1mm/1\% & 94.45 & 8.29  & 24.04 & 99.95  & $94.54 \pm 25.95$ \\
\hline
\end{tabular}%
}
\end{table}

\subsection{Baseline and Loss Function Comparison}

The undenoised $5\times10^3$ primary particle MC dose distributions serve as the lower
bound for all comparisons, achieving a mean GPR of only 26.75\% at the 2mm/2\%
criterion and 13.54\% at 1mm/1\%, confirming the poor clinical quality of
low primary particle count simulations and the necessity of a denoising step.

Training with MSE improved substantially over the raw baseline (mean GPR 92.10\%
at 2mm/2\%, 71.13\% at 1mm/1\%), but the 1mm/1\% result shows significant
residual error and a high standard deviation of 20.23\%, indicating inconsistent
performance across test cases.

Replacing MSE with the differentiable GPR loss improved both criteria markedly.
At 2mm/2\% the mean GPR rose to 99.47\% ($+7.37$), with the standard deviation
tightening from 8.01\% to 1.12\%; the gain is larger still at the stricter
1mm/1\% criterion (95.67\% vs.\ 71.13\%, $+24.54$). This confirms that directly
optimising a differentiable surrogate of the clinical metric outperforms a
pixel-wise proxy, and validates the extension of the Martinot et al.\
\cite{Martinot2023diffGammaLoss} loss to native 3D operation.

The worst-case behaviour, which matters most clinically, improves in step: the
minimum GPR rises from 43.32\% (MSE) to 85.86\% (GPR loss) at 2mm/2\%, and from
7.42\% to 34.81\% at 1mm/1\%. The GPR loss thus not only raises the mean but
sharply reduces the frequency of poor predictions.

\subsection{Pruning Results}

Structured filter pruning was applied to the GPR-trained model under two
configurations of increasing aggressiveness, with results reported after
fine-tuning.

Partial pruning of the deeper convolutional layers (conv3, conv4, conv4\_1) reduced the parameter count by 13.3\% (from 3,957,258 to 3,432,042,
15.10\,MB to 13.09\,MB) with a negligible GPR drop of 0.06\% at 2mm/2\%
(99.47\% $\rightarrow$ 99.41\%) and 0.64\% at 1mm/1\% (95.67\%
$\rightarrow$ 95.03\%) after retraining. These results confirm that the shallower branches of
each MultiScaleElaboration block carry the majority of clinically relevant
feature information, and that the deeper sequential layers (conv3--conv4\_1)
are partially redundant and can be pruned without meaningful performance
degradation.

Full-block pruning, which extended pruning to all seven convolutions per
block, achieved a 20.0\% parameter reduction (3,957,258 to 3,164,362,
15.10\,MB to 12.07\,MB). At 2mm/2\%, the mean GPR remained at 99.33\%,
only 0.14\% below the unpruned model, demonstrating effective recovery
through fine-tuning. At the stricter 1mm/1\% criterion, however, the mean
GPR dropped to 94.45\% ($-1.22\%$ from unpruned) after retraining, with a wider standard
deviation of 8.29\% compared to 6.70\%, suggesting that full-block pruning
introduces a modest but measurable loss of fine-grained spatial accuracy
that fine-tuning does not fully recover.

The greater resilience of partial pruning follows from the block structure. Each
MultiScaleElaboration block has four parallel branches of increasing depth
(conv1; conv2; conv3--conv3\_1; conv4--conv4\_1--conv4\_2). The shallower
branches, left intact under partial pruning, capture the large-scale structure
and dominant frequency content of the beam profile that are most critical for
Bragg peak localisation, whereas the deeper chains contribute mainly fine-grained
detail that is partially redundant given the encoder skip connections. Full-block
pruning also thins the shallower branches.

\subsection{Inference Time}\label{sec:inference}

Inference times were measured as forward pass only at batch size 1 on the
test set. The unpruned GPR model achieved a mean inference time of
91.53\,ms. Notably,
neither pruning configuration produced a meaningful reduction in inference
time: partial pruning increased mean inference time marginally to 95.42\,ms,
and full-block pruning to 96.37\,ms. Given the magnitude of the
reported standard deviations (up to $\pm27$\,ms), these small increases are
within run-to-run variability, and the appropriate conclusion is that pruning
left inference latency unchanged rather than that it degraded it.

A plausible explanation, consistent with known behaviour of 3D
convolutional networks at this scale, is that the forward pass is limited by
memory bandwidth and kernel launch overhead rather than by arithmetic
throughput, so that removing channels reduces FLOPs without relieving the
actual bottleneck; the reduction from 12 to 8 channels per elaboration block
may simply be too small to shift the workload into a compute-bound regime. We
emphasise, however, that we did not perform kernel-level profiling to confirm
this, and we therefore present it as a hypothesis to be verified rather than
an established finding.

The practical value of pruning should therefore be stated carefully.
Pruning did not deliver the faster inference that motivated it, and we claim no
latency benefit. What it does establish is that the architecture is
over-parameterised: up to 20\% of parameters can be removed while dose agreement
is essentially unchanged (99.33\% vs.\ 99.47\% at 2mm/2\%), with a matching 20\%
reduction in stored model size (15.10\,MB $\rightarrow$ 12.07\,MB). Beyond
storage, this yields a lighter, empirically validated backbone for follow-on
work, composite losses, higher resolutions, or larger batches, at reduced memory
cost and no measured dosimetric penalty.

\subsection{Clinical Applicability and Real-Time Feasibility}\label{sec:applicability}

A practical question is whether the end-to-end pipeline is fast enough
for adaptive or online planning. The measured 91.53\,ms forward pass per
$256\times32\times32$ subvolume is only one component: refinement also requires
generating the $5\times10^3$ primary particle input, reported at 0.65\,s per
geometry in the original study \cite{Matteo_2020_unet_IEEE}. This gives an
indicative end-to-end cost of about 0.74\,s per beam geometry against 130.30\,s
for the $10^6$ primary particle reference, a speedup of roughly two orders of
magnitude.

\subsection{Limitations}

Two limitations of the current approach are worth noting before drawing
final conclusions. First, the GPR loss by construction leaves low-dose
voxels below the cutoff threshold $\tau$ unsupervised, producing spurious
out-of-field activations that require the post-processing mask described
in Section~\ref{sec:mask}. While the mask is effective, it introduces a dependency
on the quality of the $5\times10^3$ primary particle input dose for mask
generation, which may be unreliable in very low-signal regions. Second,
the network was trained and evaluated exclusively on beams directed along
a fixed axis. Clinical treatment plans involve beams at arbitrary gantry
angles, and rotating CT and dose volumes to align with the network's
expected orientation introduces trilinear interpolation errors at
non-orthogonal angles that may degrade denoising accuracy, particularly
near bone--tissue interfaces where the Bragg peak sensitivity is highest.
Both limitations are addressed in the future work directions outlined in
Section~\ref{sec:conclusion}.
\section{Conclusion and Future Work}\label{sec:conclusion}

This study evaluated three improvements to the proton dose denoising network of
Spezialetti et al.\ \cite{Matteo_2020_unet_IEEE}: a clinically meaningful
differentiable loss, structured filter pruning, and mixed precision training.

The central finding is that replacing the MSE objective with the differentiable
GPR loss of Martinot et al.\ \cite{Martinot2023diffGammaLoss}, extended here for
the first time to native 3D operation, substantially improves clinical dose
agreement: the mean GPR rises from 92.10\% to 99.47\% at 2mm/2\% and by 24.54
percentage points at the stricter 1mm/1\% criterion, with a marked tightening of
the worst-case tail. Directly optimising a surrogate of the clinical metric thus
outperforms pixel-wise regression.

Structured pruning showed the network to be over-parameterised: up to
20\% of parameters can be removed at essentially no dosimetric cost (99.33\% vs.\
99.47\% at 2mm/2\% after fine-tuning). Pruning did not reduce inference latency,
its value is a 20\% smaller stored model and, more
importantly, an empirically validated lighter backbone for the follow-on work
below.

Taken together, these results support a concrete role for learned dose
refinement inside a treatment planning system. Planning needs dose profiles that
are both accurate and fast enough for iterative optimisation and adaptive
re-planning, which full Monte Carlo has struggled to provide. Recovering
Monte Carlo-quality profiles from a $5\times10^3$ primary particle simulation in
well under a second per geometry, rather than the roughly two minutes of the
$10^6$ primary particle reference, positions such a network as a fast dose engine
that complements rather than replaces the underlying simulation: the low-count
run retains the physical model while the network restores statistical quality.
Realising this remains subject to the integration, commissioning and quality
assurance requirements of Section~\ref{sec:applicability}.

Two directions follow from the limitations noted above. First, the out-of-field
artefact currently handled by the dilated mask could be addressed more
principledly by a composite loss combining $\mathcal{L}_{GPR}$ with a weighted
MSE term to supervise the sub-cutoff region, with careful relative normalisation
of the two terms. Second, generalising beyond the fixed beam axis to arbitrary
gantry angles, which presently requires interpolation-inducing volume rotations,
motivates a rotation-invariant architecture as a prerequisite for integration
into a full planning workflow.

\section*{Disclosure of Interests}
The authors have no competing interests to declare that are relevant to the content of this article. 
\bibliographystyle{splncs04}
\bibliography{references}
\end{document}